\documentclass[twoside,twocolumn,9pt]{article}
\usepackage{extsizes}
\usepackage[super,sort&compress,comma]{natbib} 
\usepackage[version=3]{mhchem}
\usepackage[left=1.5cm, right=1.5cm, top=1.785cm, bottom=2.0cm]{geometry}
\usepackage{balance}
\usepackage{mathptmx}
\usepackage{sectsty}
\usepackage{graphicx} 
\usepackage{lastpage}
\usepackage[format=plain,justification=justified,singlelinecheck=false,font={stretch=1.125,small,sf},labelfont=bf,labelsep=space]{caption}
\usepackage{float}
\usepackage{fancyhdr}
\usepackage{fnpos}
\usepackage[english]{babel}
\addto{\captionsenglish}{%
  \renewcommand{\refname}{Notes and references}
}
\usepackage{graphicx} 
\usepackage{array}
\usepackage{droidsans}
\usepackage{charter}
\usepackage[T1]{fontenc}
\usepackage[usenames,dvipsnames]{xcolor}
\usepackage{setspace}
\usepackage[compact]{titlesec}
\usepackage{hyperref}

\usepackage{epstopdf}

\definecolor{cream}{RGB}{222,217,201}

\begin{document}

\pagestyle{fancy}
\thispagestyle{plain}
\fancypagestyle{plain}{
\renewcommand{\headrulewidth}{0pt}
}

\makeFNbottom
\makeatletter
\renewcommand\LARGE{\@setfontsize\LARGE{15pt}{17}}
\renewcommand\Large{\@setfontsize\Large{12pt}{14}}
\renewcommand\large{\@setfontsize\large{10pt}{12}}
\renewcommand\footnotesize{\@setfontsize\footnotesize{7pt}{10}}
\makeatother

\renewcommand{\thefootnote}{\fnsymbol{footnote}}
\renewcommand\footnoterule{\vspace*{1pt}%
\color{cream}\hrule width 3.5in height 0.4pt \color{black}\vspace*{5pt}} 
\setcounter{secnumdepth}{5}

\makeatletter 
\renewcommand\@biblabel[1]{#1}            
\renewcommand\@makefntext[1]%
{\noindent\makebox[0pt][r]{\@thefnmark\,}#1}
\makeatother 
\renewcommand{\figurename}{\small{Fig.}~}
\sectionfont{\sffamily\Large}
\subsectionfont{\normalsize}
\subsubsectionfont{\bf}
\setstretch{1.125} 
\setlength{\skip\footins}{0.8cm}
\setlength{\footnotesep}{0.25cm}
\setlength{\jot}{10pt}
\titlespacing*{\section}{0pt}{4pt}{4pt}
\titlespacing*{\subsection}{0pt}{15pt}{1pt}

\fancyfoot{}
\fancyfoot[LO,RE]{\vspace{-7.1pt}\includegraphics[height=9pt]{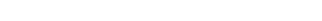}}
\fancyfoot[CO]{\vspace{-7.1pt}\hspace{11.9cm}\includegraphics{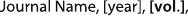}}
\fancyfoot[CE]{\vspace{-7.2pt}\hspace{-13.2cm}\includegraphics{head_foot/RF}}
\fancyfoot[RO]{\footnotesize{\sffamily{1--\pageref{LastPage} ~\textbar  \hspace{2pt}\thepage}}}
\fancyfoot[LE]{\footnotesize{\sffamily{\thepage~\textbar\hspace{4.65cm} 1--\pageref{LastPage}}}}
\fancyhead{}
\renewcommand{\headrulewidth}{0pt} 
\renewcommand{\footrulewidth}{0pt}
\setlength{\arrayrulewidth}{1pt}
\setlength{\columnsep}{6.5mm}
\setlength\bibsep{1pt}

\makeatletter 
\newlength{\figrulesep} 
\setlength{\figrulesep}{0.5\textfloatsep} 

\newcommand{\topfigrule}{\vspace*{-1pt}%
\noindent{\color{cream}\rule[-\figrulesep]{\columnwidth}{1.5pt}} }

\newcommand{\botfigrule}{\vspace*{-2pt}%
\noindent{\color{cream}\rule[\figrulesep]{\columnwidth}{1.5pt}} }

\newcommand{\dblfigrule}{\vspace*{-1pt}%
\noindent{\color{cream}\rule[-\figrulesep]{\textwidth}{1.5pt}} }

\makeatother

\twocolumn[
  \begin{@twocolumnfalse}
{\includegraphics[height=30pt]{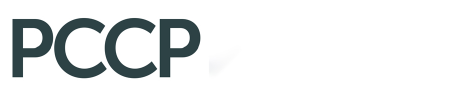}\hfill\raisebox{0pt}[0pt][0pt]{\includegraphics[height=55pt]{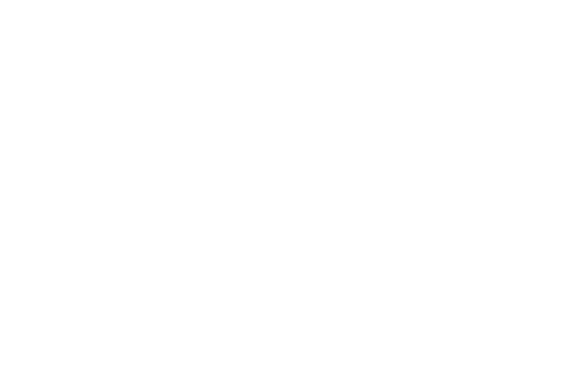}}\\begin{equation}1ex]
\includegraphics[width=18.5cm]{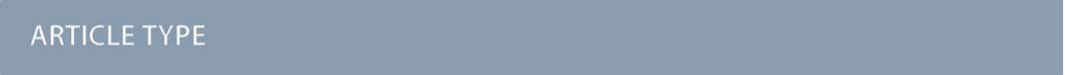}}\par
\vspace{1em}
\sffamily
\begin{tabular}{m{4.5cm} p{13.5cm} }

\includegraphics{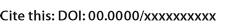} & \noindent\LARGE{\textbf{Quantum Entanglement Control in Two-Spin-1/2 NMR Systems Through Magnetic Fields and Temperature}} \\
\vspace{0.3cm} & \vspace{0.3cm} \\

 & \noindent\large{Fatemeh Khashami,$^{\ast}$\textit{$^{a}$} and Stefan  Gl\"oggler\textit{$^{a}$}} \\

\includegraphics{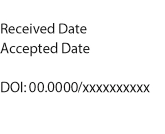} & \noindent\normalsize{We investigate quantum entanglement in two-spin-1/2 NMR systems at thermal equilibrium under external magnetic fields. We derive closed-form analytical expressions for the entanglement of the system and show how the entanglement depends on temperature and magnetic field strength, resulting in a threshold temperature beyond which entanglement vanishes. We demonstrate that at zero temperature, the system exhibits a quantum critical point, characterized by non-analytic behavior in the measure of entanglement. We further develop analytical criterion for level crossing, which serves as a condition for identifying quantum critical points in both homonuclear and heteronuclear systems, and apply it to multiple settings to analyze their quantum critical points. We establish a direct link between the quantum entanglement quantifier and experimentally accessible NMR observables, enabling entanglement to be quantified through NMR signal processing. This provides a practical framework for characterizing quantum correlations using standard NMR experiments. These findings provide insights into the thermal control of quantum features, with implications for quantum-enhanced NMR, low-temperature spectroscopy, and emerging quantum technologies.} 
\\

\end{tabular}

 \end{@twocolumnfalse} \vspace{0.6cm}

  ]

\renewcommand*\rmdefault{bch}\normalfont\upshape
\rmfamily
\section*{}
\vspace{-1cm}


\footnotetext{\textit{$^{a}$~Advanced Imaging Research Center, University of Texas Southwestern Medical Center,  Dallas, Texas, USA. E-mail: fatemeh.khashami@utsouthwestern.edu}}



\section{Introduction}

The structure of two interacting spins in thermal equilibrium under an external magnetic field gives rise to intriguing phenomena, particularly in nuclear magnetic resonance (NMR) spectroscopy \cite{slichter2013principles,bloch1953principle}. The degeneracy of such a two-spin-1/2 system is a crucial parameter, described by the expression $I = (2I_1 + 1)(2I_2 + 1)$, where $I_1$ and $I_2$ denote the spin quantum numbers for spin-$1$ and spin-$2$, respectively. The resulting NMR spectrum displays four distinct peaks for two-spin-1/2 systems characterized by four degenerate energy states \cite{ernst1987principles,wootters1998entanglement}.

Within the context of two-spin-1/2 systems, studies have explored intricate quantum coherence and entanglement within atomic systems, demonstrating fundamental insights into hyperfine structures \cite{maleki2021naturalennett,tommasini1997hydrogen,harilal2020hyperfine}. Beyond hyperfine structures, entanglement in spin systems has been investigated in various contexts, particularly quantum information processing applications \cite{furman2009nuclear,yamamoto2007feedback,satoori2022entanglement,maleki2015entanglement,broekhoven2024protocol}. More specifically, considerable attention has been devoted to quantum entanglement in different Heisenberg spin models from the perspective of quantum information theory (see \cite{arnesen2001natural,wang2001entanglement,guo2003thermal,asoudeh2005thermal,vcenvcarikova2020unconventional,adamyan2020quantum} and references therein).

Thermal polarization, arising from the statistical distribution of spin states, introduces an additional feature of non-classical behavior in quantum systems \cite{herzog2014boundary,schmidt2014using}. Investigating quantum entanglement in thermally polarized NMR systems provides deeper insights into their quantum nature \cite{hovav2013theoretical,khashami2023fundamentals}.
Such investigations are crucial not only for understanding NMR spectroscopy in chemical systems \cite{fan2016applications,bloch1946nuclear} but also for advancing entanglement-assisted quantum technologies and quantum information processing \cite{nielsen2010quantum,horodecki2009quantum,maleki2021quantum,jones2001nmr,oliveira2011nmr,maleki2016generation}.

Recent studies have extensively examined scalar-coupled spin-1/2 systems in NMR, investigating both homonuclear and heteronuclear configurations and providing valuable insight into their spectral behavior under varying coupling strengths and magnetic field conditions~\cite{donovan2014heteronuclear,vuichoud2015measuring,appelt2010paths,wang2001entanglement}. These investigations have primarily focused on spectral features and spin dynamics, advancing our understanding of coherence, relaxation, and signal evolution in coupled spin systems. However, the quantum mechanical aspects of entanglement in these systems, particularly its thermal behavior and dependence on field strength in the context of quantum chemistry, needs further attentions ~\cite{jones2011quantum,cory2000nmr}.

In this study, we consider two-spin-1/2 systems in thermal equilibrium subjected to an external magnetic field. We investigate entanglement and its dependence on parameters such as temperature and magnetic field strength \cite{arute2019quantum,donovan2014heteronuclear}. Temperature plays a crucial role in determining whether quantum correlations persist, as thermal fluctuations tend to destroy entanglement. The quantification of entanglement is carried out using the concurrence measure. Notably, we identify a threshold temperature beyond which entanglement vanishes, regardless of the strength of the applied magnetic field. At zero temperature, the behavior of the entanglement exhibits non-analytic features that signal a quantum phase transition
\cite{vojta2003quantum,chaboussant1998nuclear,wang2001entanglement}.

Furthermore, we explore the transition between strong and weak coupling regimes facilitated by the magnetic field. This transition is marked by a significant feature known as the crossing point, where the energy levels of the system intersect as the system evolves from low magnetic fields to high magnetic fields. 
The results of this investigation highlight how the cooperative effects of external magnetic fields, spin coupling mechanisms, and temperature profoundly affect the entanglement of an NMR system. These insights not only hold theoretical importance but also pave the way for practical applications in NMR technology and quantum information processes. By analyzing the two-spin-1/2 systems, we aim to deepen the understanding of quantum properties of NMR systems at thermal equilibrium under external magnetic fields.

\section{Methods}

\subsection{Hamiltonian Formalism of the Two-Spin-1/2 NMR System}

The system we examine here consists of two interacting spins, where magnetic dipole moments lead to spin-spin coupling. These interactions contribute to the total Hamiltonian of the system \cite{khashami2023fundamentals, mamone2020singlet, donovan2014heteronuclear,rudowicz2001spin}. The Hamiltonian operator for a two-spin-1/2 system is expressed as
\begin{equation}
\mathcal{H} = \mathcal{H}_z + \mathcal{H}_{\boldsymbol{J}},
\end{equation}
where $\mathcal{H}_z$ describes the independent (non-interacting) Hamiltonian associated with the Zeeman energy levels and $\mathcal{H}_{\boldsymbol{J}}$ accounts for the spin-spin interaction term. Using Pauli matrices (with $\hbar=1$), the full Hamiltonian can be expressed as
\begin{equation}
\mathcal{H} = \frac{1}{2}(\omega_{\Sigma} + \omega_{\delta}) I_{1z} + \frac{1}{2}(\omega_{\Sigma} - \omega_{\delta}) I_{2z} + 
\boldsymbol{J}\, {I}_1 \cdot {I}_2, \label{Hamiltonian_G}
\end{equation}
where ${\boldsymbol{J}}$ is the coupling constant in ${\text{rad}}/{\text{sec}}$ \cite{vuichoud2015measuring, ivanov2022chemically}.
The sum and difference of the Larmor frequencies are defined as $\omega_{\Sigma} = \omega_1 + \omega_2$ and $\omega_{\delta} = \omega_1 - \omega_2$. Spin operators follow $I_{i\alpha} = \sigma_{i\alpha}/2$, where $\sigma_{i\alpha}$ are Pauli matrices.
The Larmor frequency of each spin is $\omega_i = \omega_0 (1 - \delta_i) - \omega_{\text{RF}}$ for spin-1 and spin-2, 
where $\delta_i$ is the chemical shift, $\omega_0 = \gamma_i B_0$ is the Larmor frequency and $\omega_{\text{RF}}$ is the applied radio frequency pulse (RF) \cite{hore2015nmr,schaublin1974fourier}.
The last term in Eq.~(\ref{Hamiltonian_G}) reflects the interaction between the two spins, directly affecting the system's energy levels. Therefore, the Hamiltonian can also be represented in matrix form as \cite{khashami2023fundamentals,eykyn2021extended}
\begin{equation}
\mathcal{H} = \frac{1}{2}
\begin{bmatrix}
+\omega_{\Sigma} + \frac{1}{2}\boldsymbol{J} & 0 & 0 & 0 \\
0 & +\omega_{\delta} - \frac{1}{2}\boldsymbol{J} & \boldsymbol{J} & 0 \\
0 & \boldsymbol{J} & -\omega_{\delta} - \frac{1}{2}\boldsymbol{J} & 0 \\
0 & 0 & 0 & -\omega_{\Sigma} + \frac{1}{2}\boldsymbol{J}
\end{bmatrix}.
\end{equation}

\begin{figure}[h]
\centering
  \includegraphics[width=\columnwidth]{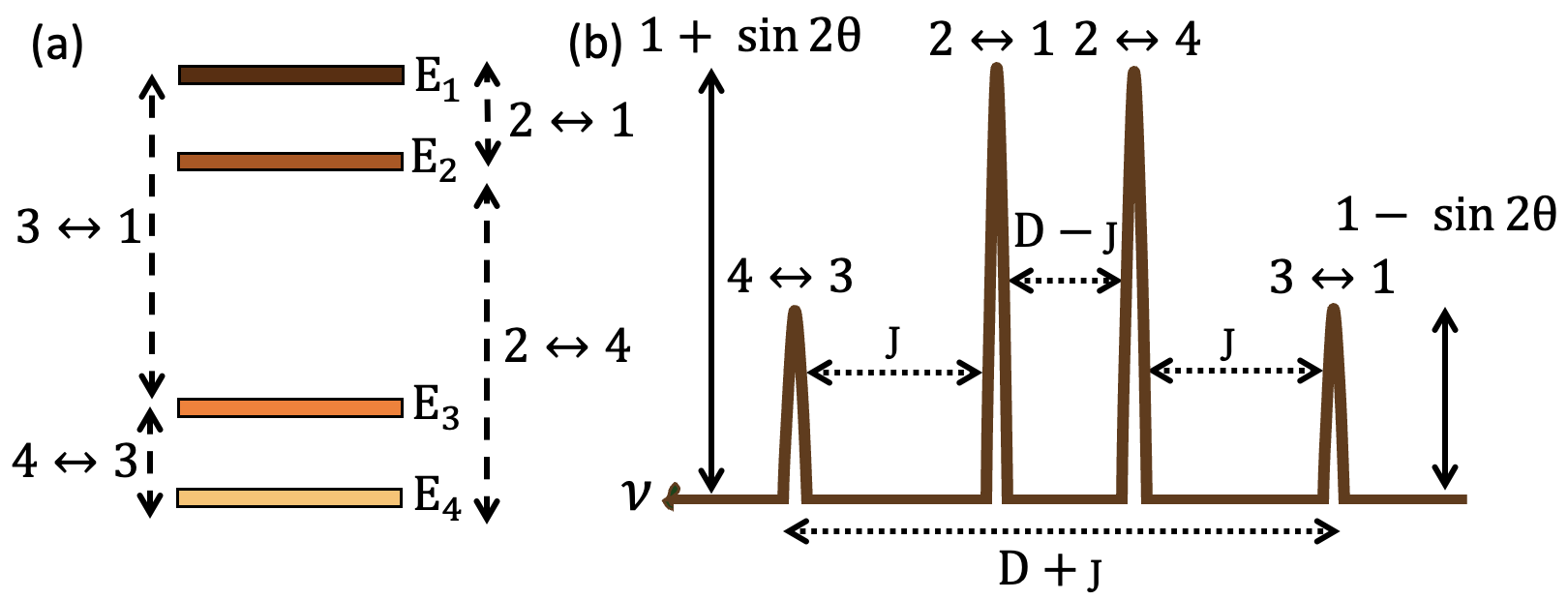}
\caption{(a) Zeeman energy levels of a two-spin-1/2 system, with energy values $\mathrm{E}_1$, $\mathrm{E}_2$, $\mathrm{E}_3$, and $\mathrm{E}_4$. (b) Corresponding NMR spectrum, where frequency increases from right to left. The frequency differences are represented by dashed lines. Signal intensities follow the ''roofing effect'', given by $1 \pm \sin 2\theta$.}
  \label{Fig1_evergyLevel}
\end{figure}

Also, the system Hamiltonian can be written in the basis of the coupled spin states ${|\phi_1\rangle, |\phi_2\rangle, |\phi_3\rangle, |\phi_4\rangle}$ as
\begin{equation}
\mathcal{H} = \mathrm{E}_1 |\phi_1\rangle \langle\phi_1| + \mathrm{E}_2 |\phi_2\rangle \langle\phi_2| + \mathrm{E}_3 |\phi_3\rangle \langle\phi_3| + \mathrm{E}_4 |\phi_4\rangle \langle\phi_4|, \label{HamiltonianTotal}
\end{equation}
where the basis of a two-spin-1/2 system is
\begin{equation} 
\begin{aligned}\label{eigenstates}
{|\phi_1\rangle } & = |\alpha\alpha\rangle, \quad\quad
{|\phi_2\rangle }  = \cos\theta |\alpha\beta\rangle +\sin\theta | \beta\alpha\rangle
, \\
{|\phi_3\rangle }&= -\sin\theta |\alpha\beta\rangle +\cos\theta |\beta\alpha\rangle, \quad \quad
{|\phi_4\rangle } = |\beta\beta\rangle ,
\end{aligned}    
\end{equation}
with the mixing angle $\theta$ defined as $\sin 2 \theta={\boldsymbol{J}}/{D}$ and $\tan 2 \theta={\boldsymbol{J}}/{\omega_{\delta}}$
with the coupling constant $D=\sqrt{\omega_{\delta}^2+\boldsymbol{J}^2}$. 
The energy eigenvalues of the system, corresponding to the energy levels  Fig. \ref{Fig1_evergyLevel}(a), are given by
\begin{equation}
 \begin{aligned}
&\mathrm{E}_1=\frac{1}{2} (\omega_{\Sigma}+\frac{1}{2} \boldsymbol{J}),  \quad    
\mathrm{E}_2=\frac{1}{2}(D-\frac{1}{2} \boldsymbol{J}), \quad 
 \\
&\mathrm{E}_3=-\frac{1}{2} (D+\frac{1}{2} \boldsymbol{J}), \quad 
\mathrm{E}_4=\frac{1}{2} (-\omega_{\Sigma}+\frac{1}{2} \boldsymbol{J}). 
\end{aligned} \label{EnergyLevel}  
\end{equation}

The NMR spectrum of a two-spin-1/2 system contains four peaks, as shown in Fig.~\ref{Fig1_evergyLevel}(b). 
The transition frequencies, given by $\Delta \nu = \Delta E / \gamma$, follow a characteristic pattern determined by the energy level spacings. The highest frequency corresponds to the $4 \leftrightarrow 3$ transition, followed by $2 \leftrightarrow 1$ and $2 \leftrightarrow 4$, with the $3 \leftrightarrow 1$ transition appearing at the lowest frequency.
The frequency difference between the transitions ${2 \leftrightarrow 4}$ and ${2 \leftrightarrow 1}$ is given by $D - \boldsymbol{J}$, and the difference between ${2 \leftrightarrow 4}$ and ${3 \leftrightarrow 1}$ is $\boldsymbol{J}$. 
Moreover, the ''roofing effect'' represented by the factor $1 \pm \sin 2\theta$ leads to enhanced intensity of the inner peaks compared to the outer ones, resulting from the asymmetric distribution of transition energies \cite{vuichoud2015measuring,khashami2023fundamentals}.

\subsection{Thermal State Representation of the Two-Spin-1/2 NMR System}

In a two-spin-1/2 system, the density matrix in its diagonal form can be expressed as
\begin{equation}
 \rho = p_{1} |\phi_1\rangle \langle\phi_1| + p_{2} |\phi_2\rangle \langle\phi_2| + p_{3} |\phi_3\rangle \langle\phi_3| + p_{4} |\phi_4\rangle \langle\phi_4|,   
\end{equation}
where the coefficients $p_1$, $p_2$, $p_3$, and $p_4$ represent the occupation probabilities of the corresponding quantum states.

More generally, for $n$ eigenstates, the density matrix can be written in compact summation form as
\begin{equation}
\rho = \sum_{i=1}^n p_i \left|\phi_i\right\rangle \left\langle\phi_i\right|, \label{densityMatrix_G}
\end{equation}
where the probabilities satisfy the normalization condition $\sum_i p_i = 1$ and each $p_i \geq 0$, ensuring that $\operatorname{Tr}\rho = 1$.
The probability of the system occupying the state $\left|\phi_i\right\rangle$ at thermal equilibrium follows the Boltzmann distribution $p_i = e^{-\beta \mathrm{E}_i}/Z$,
where $Z$ denotes the partition function of the system. The inverse temperature is given by $\beta = {1}/{k_B T}$, where $k_B = 1.38065\times 10^{-23}$ \text{Joule/Kelvin} is the Boltzmann constant, and $T$ denotes the temperature in Kelvin (K).

At thermal equilibrium, the thermal density matrix for a two-spin-1/2 system is ${\rho} = \sum_{i=1}^4 e^{-\beta \mathrm{E}_i} |\phi_i\rangle\langle\phi_i|/Z$.
Using the four energy states given in Eq. (\ref{EnergyLevel}), the partition function for this system is expressed as $Z= \sum_{i=1}^4 e^{-\beta E_i}$, and can be explicitly expressed as
\begin{equation}
 Z=2  e^{ \frac{\beta \boldsymbol{J}}{4}}(e^{-\frac{\beta \boldsymbol{J}}{2}} \cosh (\frac{\beta \omega_{\Sigma}}{2})+\cosh (\frac{\beta D}{2})). \label{partitionFunction}  
\end{equation}

Furthermore, the density matrix can also be directly expressed in the computational basis $\{|\alpha\alpha\rangle, |\alpha\beta\rangle, |\beta\alpha\rangle, |\beta\beta\rangle \}$ as
\begin{equation}\label{density}
\rho = 
\begin{pmatrix}
\rho_{11} & 0 & 0 & 0 \\
0 & \rho_{22} & \rho_{23} & 0 \\
0 & \rho_{32} & \rho_{33} & 0 \\
0 & 0 & 0 & \rho_{44}
\end{pmatrix}, 
\end{equation}
where the elements of the density matrix are given by
\begin{align}
\rho_{11} &= \frac{1}{Z} e^{-\beta E_1}, \quad
\rho_{22}  = \frac{1}{Z} (e^{-\beta E_2} \cos^2\theta + e^{-\beta E_3} \sin^2\theta),\nonumber\\
\rho_{33} &= \frac{1}{Z} (e^{-\beta E_2} \sin^2\theta + e^{-\beta E_3} \cos^2\theta), \quad \rho_{44} = \frac{1}{Z} e^{-\beta E_4},\nonumber \\
\rho_{23} &= \rho_{32} = \frac{1}{Z} (e^{-\beta E_2} - e^{-\beta E_3}) \sin\theta \cos\theta. 
\label{densityMatrixElement}
\end{align}

The density matrix above gives the thermal state of a two-spin-1/2 system under an external magnetic field. With this explicit form of the developed density matrix, we analyze the quantum entanglement of the system that characterizes the non-classical features of the system in the following section. We use this density matrix to derive closed-form analytical expressions for the concurrence, which quantify the degree of entanglement.

\section{Results}

\subsection{ Quantifying  Entanglement in the Two-Spin-1/2 NMR System}

For a two-spin-1/2 system, the entanglement can be evaluated using the concurrence \cite{wootters1998entanglement,hill1997entanglement} 
\begin{equation}
{C} = \mathrm{max}\{0, \lambda_1 - \lambda_2 - \lambda_3 - \lambda_4\},
\end{equation}
where $\lambda_i$ denote the eigenvalues of the Hermitian matrix $R = \sqrt{\sqrt{\rho}\, \tilde{\rho}\, \sqrt{\rho}}$, such that $\lambda_1 \geq \lambda_2 \geq \lambda_3 \geq \lambda_4$. In addition, $\tilde{\rho}$ is the spin-flipped density matrix, defined as $\tilde{\rho} = (\sigma_y \otimes \sigma_y)\, \rho^*\, (\sigma_y \otimes \sigma_y)$
with $\rho^*$ denoting the complex conjugate of $\rho$, and $\sigma_y$ is the Pauli matrix corresponding to the $y$ component of the spin operator. A maximally entangled quantum state corresponds to
$C=1$, whereas a fully separable state results in
$C=0$.

For the thermal density matrix of the two-spin-1/2 NMR system, we obtain the concurrence  of the system as
\begin{equation}
{C} = \max\{0, |p_2 - p_3| |\sin 2\theta| - 2 \sqrt{p_1 p_4}\}, \label{concurrence_G}
\end{equation}
where the term $|p_2 - p_3|$ reflects the population difference between specific states and is given by
\begin{equation}
\begin{aligned}
|p_2 - p_3| &= \frac{1}{Z}|e^{-\beta E_2} - e^{-\beta E_3}| 
= \frac{2}{Z} \sinh(\frac{\beta D}{2}) e^{\frac{\beta \boldsymbol{J}}{4}},
\end{aligned}
\end{equation}
also, the term $\sqrt{p_1p_4}$ corresponds to a product of population terms and can be expressed as $\sqrt{p_1 p_4} = {e^{-\frac{\beta \boldsymbol{J}}{4}}}/Z.$
Thus, combining these expressions, the concurrence is given by
\begin{equation}
{C} = \frac{2}{Z}\, \max\left\{0, e^{\frac{\beta \boldsymbol{J}}{4}} \sinh(\frac{\beta D}{2}) \sin 2\theta - e^{-\frac{\beta \boldsymbol{J}}{4}}\right\}, \label{concurrence}
\end{equation}
where, without loss of generality, we assume $J\geq0$, and hence $\sin 2 \theta$ gives a non-negative number.
This relation demonstrates that the entanglement properties of an NMR system depend on several factors, including $T$, $\boldsymbol{J}$, {D}, and $\omega_{\Sigma}$.
By substituting the expression for the partition function from Eq.~(\ref{partitionFunction}) into Eq.~(\ref{concurrence}), the concurrence can be expressed as
\begin{equation}
{C} = \max\left\{0, \frac{\sinh(\frac{\beta D}{2})\sin 2\theta - e^{-\frac{\beta \boldsymbol{J}}{2}}}{e^{-\frac{\beta \boldsymbol{J}}{2}}\cosh(\frac{\beta \omega_{\Sigma}}{2}) + \cosh(\frac{\beta D}{2})}\right\}. \label{concurrence_General}
\end{equation}

This expression can be evaluated numerically for a specific parameter for an NMR system, as illustrated in Fig.~\ref{Fig2_concurrence}. 
For an entangled state, the non-zero concurrence satisfies ${C} > 0$, which imposes the condition as 
\begin{equation}
\sinh(\frac{\beta D}{2}) \sin 2\theta > e^{-\frac{\beta \boldsymbol{J}}{2}},
\end{equation}
substituting $\sin 2\theta = {\boldsymbol{J}}/{D}$ into this expression, the  disappearance point of the entanglement reduces to the following nonlinear equation 
\begin{equation}\label{nonlinear_critria}
\boldsymbol{J}\, e^{\frac{\beta \boldsymbol{J}}{2}} = \frac{D}{\sinh\left({\beta D}/{2}\right)}.
\end{equation}
This equation provides the threshold temperature above which the entanglement vanishes. 

\subsubsection{Entanglement in Weak Spin-Spin Coupling} \label{weakB}

In the weak coupling regime, where $\boldsymbol{J}$ is much smaller than $ \omega_{\delta} $, the mixing angle remains small. Under this condition, $ D $ can be approximated in terms of $ \omega_{\delta} $ with a small correction due to the coupling strength \cite{khashami2023fundamentals}.
By setting $\boldsymbol{J} \simeq 0$, Eq.~(\ref{EnergyLevel}) gives the energy eigenvalues  $\mathrm{E}_1 = - \mathrm{E}_4 = {\omega_{\Sigma}}/{2}$, and $\mathrm{E}_2 = - \mathrm{E}_3 = {\omega_{\delta}}/{2}$,
where the energy ordering follows $\mathrm{E}_1 \geq \mathrm{E}_2 \geq \mathrm{E}_3 \geq \mathrm{E}_4$, indicating that the Zeeman term in Eq.~(\ref{Hamiltonian_G}) dominates over the spin-spin coupling. The eigenstates for weak coupling, consistent with Eq.~(\ref{eigenstates}), are  
\begin{equation}
\begin{aligned}
&|\phi_1\rangle = |\alpha\alpha \rangle, \quad
|\phi_2\rangle = |\alpha\beta\rangle, \quad
|\phi_3\rangle = |\beta\alpha\rangle, \quad
|\phi_4\rangle = |\beta\beta\rangle. \label{statedd}
\end{aligned}
\end{equation}

Since $\sin2\theta \approx 0$, all the peaks have the same intensity, as can be deduced from Fig.~\ref{Fig1_evergyLevel}(b).   
The density matrix for the weakly coupled system, based on Eq.~(\ref{densityMatrixElement}), is nearly diagonal. The off-diagonal coherence term is strongly suppressed due to the dominance of the Zeeman term. This suppression prevents the coherent superposition between $|\beta\alpha\rangle$ and $|\alpha\beta\rangle$. Consequently, entanglement is essentially absent. Therefore, the density matrix in Eq. (\ref{density}) is written based on population terms, exhibiting only classical statistics, with negligible entanglement. Each spin transitions independently, with the observed uniformity of the NMR spectrum.

\subsubsection{Entanglement in Strong Spin-Spin Coupling} \label{strB}

In the strong coupling regime, where $\boldsymbol{J} $ is much larger than $\omega_{\delta} $, the expression for $D $ can be approximated, mainly depending on $\boldsymbol{J} $ with a small correction from $\omega_{\delta} $ \cite{khashami2023fundamentals}. This approximation results in a frequency difference between the transitions ${4 \leftrightarrow 3} $ and ${2 \leftrightarrow 1} $, as well as ${2 \leftrightarrow 4} $ and ${3 \leftrightarrow 1} $, both equal to $\boldsymbol{J} $.
The eigenstates characterized by $\tan 2 \theta \gg 1$ (where $\theta = \pi/4$), are organized as a triplet state (for ${I}=1$ identified by angular number $m_I=0,\pm1$) by 
\begin{equation}
\begin{aligned}
|\phi_1\rangle = |\alpha\alpha\rangle,  \quad
|\phi_2\rangle = \frac{1}{\sqrt{2}}(|\alpha\beta \rangle + |\beta\alpha\rangle),\quad
|\phi_4\rangle = | \beta\beta\rangle,\label{angular_T}
\end{aligned} 
\end{equation}
and a singlet state (for ${I}=0$ identified by angular number $m_I=0$) as 
\begin{equation}
\left|\phi_3\right\rangle=\frac{1}{\sqrt{2}}(|\beta\alpha\rangle-|\alpha\beta\rangle).  \label{angular_S} 
\end{equation}

Based on selection rules from quantum mechanics, transitions between the singlet and triplet states are forbidden due to the difference in total spin quantum number $I$. As a result, these transitions appear to be weak or completely absent in the spectrum. 
In this setting, highly entangled states can be realized, more suitable for quantum information applications.

\subsubsection{Temperature-Dependent Entanglement in Homonuclear and Heteronuclear Spin Systems}

\begin{figure}[h]
\centering
\includegraphics[width=\columnwidth]{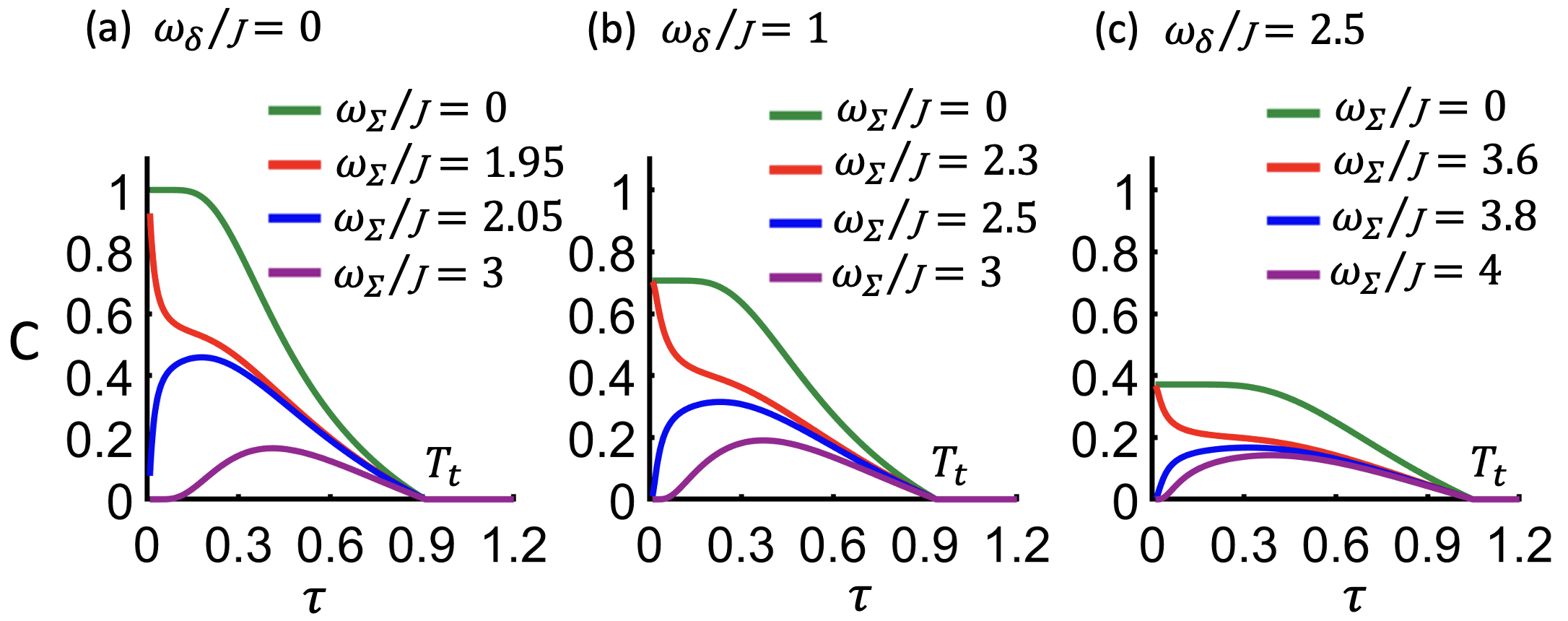}
    \caption{Concurrence as a function of temperature for different dimensionless ratio $\omega_{\Sigma}/\boldsymbol{J}$. Panels (a), (b), and (c) correspond to $ \omega_{\delta}/\boldsymbol{J} = 0 $ (a homonuclear system), $ \omega_{\delta}/\boldsymbol{J} = 1$ (a heteronuclear system), and $ \omega_{\delta}/\boldsymbol{J} = 2.5$ (a heteronuclear system), respectively.
    The re-scaled temperature parameter is $\tau = k_B T /\boldsymbol{J}$ in Eq.~(\ref{concurrence_General}), and the threshold temperature is $T_t$.}
\label{Fig2_concurrence}
\end{figure}

Temperature plays a fundamental role in the entanglement properties of two-spin-1/2 systems \cite{wang2001entanglement,maleki2021naturalennett}. 
In the case of a homonuclear system, the angular frequencies are equal, i.e., $\omega_1 = \omega_2 = \omega$; this results in $\omega_\delta = 0$ and $\omega_\Sigma = 2\omega$. Under these conditions, the system parameter simplifies to $D = \boldsymbol{J}$, thus the expression for concurrence simplifies to \cite{arnesen2001natural,wang2001entanglement,nielsen2000quantum}
\begin{equation}
{C} = \max\left\{0, \frac{e^{\beta \boldsymbol{J}} - 3}{2 \cosh(\beta \omega) + e^{\beta \boldsymbol{J}} + 1}\right\}. \label{concurrence_simple}
\end{equation}
To satisfy ${C} > 0$, the term $e^{\beta \boldsymbol{J}} - 3$ must be positive, leading to the condition $\beta>{\ln 3}/{\boldsymbol{J}}$ or equivalently $k_B T<{\boldsymbol{J}}/{\ln 3}$. This establishes the threshold temperature, $T_t$, above which entanglement vanishes
\begin{equation}
T_t = \frac{\boldsymbol{J}}{k_B \ln 3}.
\end{equation}

While the general threshold temperature for entanglement of the density matrix in Eq. (\ref{density}) can be determined for the 
condition in Eq.~(\ref{nonlinear_critria}), the simplified threshold temperature for homonuclear systems, $T_t$, coincides with the entanglement condition found in various contexts, including Heisenberg models \cite{arnesen2001natural,nielsen2000quantum} and hyperfine-structure states in atomic hydrogen \cite{maleki2021naturalennett}. However, inhomogeneity in the magnetic field can modify this condition ~\cite{asoudeh2005thermal}. 
The threshold temperature can also be influenced by variations in the anisotropy parameter, leading to modified entanglement conditions in these systems \cite{wang2001entanglement,guo2003thermal}.

Figure~\ref{Fig2_concurrence}(a) illustrates the concurrence as a function of re-scaled temperature $\tau = k_B T /\boldsymbol{J}$ for a homonuclear system under various values of $\omega_{\Sigma}/\boldsymbol{J}$. At low temperatures, concurrence remains high when $\omega_{\Sigma}/\boldsymbol{J}$ is small, indicating strong entanglement under weak magnetic fields. As $\omega_{\Sigma}/\boldsymbol{J}$ increases, the term $\cosh(\beta \omega)$ contributes more significantly in Eq.~(\ref{concurrence_simple}), leading to a rapid suppression of entanglement. Increasing temperature causes the concurrence to decline gradually, eventually reaching zero at  $T_t$. The lower values of $\omega_{\Sigma}/\boldsymbol{J}$ extend the temperature range in which the entanglement survives, whereas the higher values lead to a faster disappearance of the entanglement. These findings emphasize that temperature is not just a background condition but a controllable parameter that governs the entanglement in spin systems.
In contrast, heteronuclear systems, where $\omega_1 \neq \omega_2$, display a different thermal response \cite{appelt2010paths}. Here, $\omega_\delta$ introduce asymmetry due to the distinct gyromagnetic ratios of the two nuclei. 
Figures~\ref{Fig2_concurrence}(b)-(c) illustrate this behavior, showing that even at low temperatures, concurrence is significantly lower than in the homonuclear case. Furthermore, entanglement decreases gradually and vanishes at higher values of $\omega_{\Sigma}/\boldsymbol{J}$.

\begin{table}[h!]
\centering
\resizebox{\columnwidth}{!}{%
\begin{tabular}{|l|r|r|}
\hline 
{System} & $\boldsymbol{J}/2\pi$ (Hz) & {$T_t$ \quad} \\
\hline
Ethanol & 7 & 0.31 nK \\
Complex Organic Molecules & 150 & 6.5 nK \\
Platinum--Phosphorus Bonding & 3096 & 0.14 \textmu K \\
Phosphorus-31, platinum-195 and lead-207 & 14500 & 0.63 \textmu K \\
Phosphorus-31, platinum-195 and lead-207 & 18500 & 0.81 \textmu K \\
Hyperfine Interaction & 1.4$\times$$10^9$ & 61.2 mK \\
Metal-Carboxylate & 1.6$\times$$10^{13}$ & 680.9 K \\
\hline
\end{tabular}%
}
\caption{Scalar couplings and their corresponding temperatures for various systems} 
\label{table1}
\end{table}

Moreover, the derived threshold temperature, $k_B T_t = \hbar \boldsymbol{J} / \ln{3}$, provides a direct relationship between the spin-spin interaction strength and the thermal robustness of quantum correlations~\cite{guo2010entanglement,wang2001entanglement}. Table~\ref{table1} summarizes this relation for a range of systems. For instance, scalar couplings in organic molecules typically fall within the range of a few to $150$ Hz, corresponding to $T_t$ on the order of nK \cite{soltani2016entanglement,koenig2005lipid}. More notably, $\boldsymbol{J}$-couplings involving heavy elements can be significantly stronger. For example, the scalar coupling constant for platinum–phosphorus bonding is on the order of a few kHz, yielding $T_t$ in the lower \textmu K range \cite{chatt1975unusually}. In another case, couplings between Phosphorus-$31$, platinum-$195$, and lead-$207$ nuclei span a broader range in the kHz regime, corresponding to slightly higher $T_t$ \cite{carr1982phosphorus}. These examples highlight the wide range of $\boldsymbol{J}$-coupling strengths, making them a particularly valuable focus for studying thermal energy scales. Similar energy scales can also arise from other types of interaction. In atomic hydrogen, hyperfine-level coupling strengths are on the order of $1.4$\,GHz, which results in a $T_t$ of about $60$ mK \cite{maleki2021naturalennett}. In extreme cases, such as syn-syn metal-carboxylate bonding in metal-organic complexes, the coupling strengths may exceed $15{,}600$\,GHz, resulting in $T_t$ of the order of hundreds of kelvin \cite{souza2009entanglement,cruz2022quantum}.  Likewise, dipolar couplings observed in organic molecules or solid-state NMR systems often reach several kHz, corresponding to \textmu K or even up to the mK range in some strongly coupled systems \cite{ji2021recent,wu1994high}.

\subsubsection{Frequency-Driven Entanglement in Homonuclear and Heteronuclear Spin Systems}

\begin{figure}[h]
\centering
\includegraphics[width=\columnwidth]{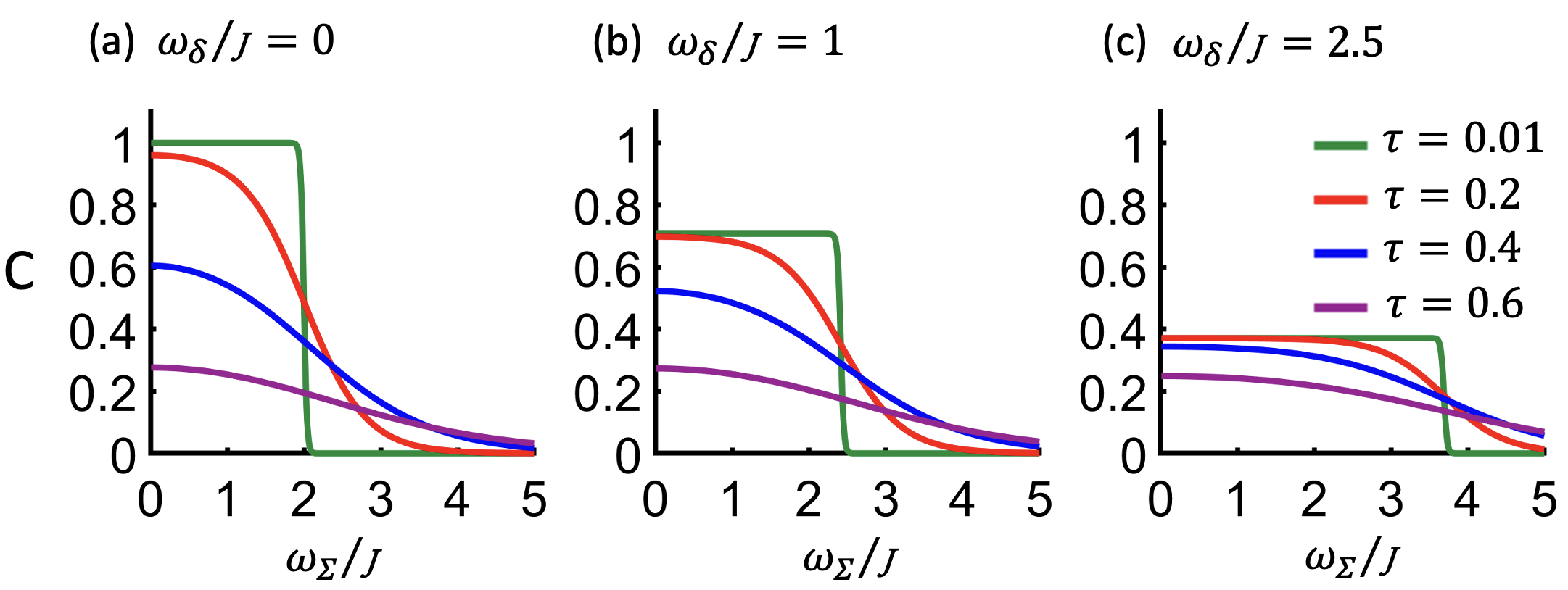}
    \caption{Concurrence as a function of the normalized frequency ratio $ \omega_{\Sigma}/\boldsymbol{J}$ for varying re-scaled temperatures. Panels (a), (b), and (c) correspond to $ \omega_{\delta}/\boldsymbol{J} = 0$ (a homonuclear system), $ \omega_{\delta}/\boldsymbol{J} = 1$ (a heteronuclear system), and $ \omega_{\delta}/\boldsymbol{J} = 2.5$ (a heteronuclear system), respectively, demonstrating how temperature influences concurrence. 
    }
\label{Fig3_concurrence}
\end{figure}

The behavior of entanglement as a function of the rescaled frequency  $\omega_{\Sigma}/\boldsymbol{J}$ is shown in Fig.~\ref{Fig3_concurrence} for both homonuclear and heteronuclear systems at varying temperatures.
For the homonuclear case shown in Fig.~\ref{Fig3_concurrence}(a), entanglement remains robust at low temperatures, indicating strong quantum correlations. As temperature increases, the concurrence begins to decline gradually, ultimately vanishing due to thermal decoherence. This highlights the joint influence of the magnetic field and temperature in suppressing quantum entanglement.
Figures~\ref{Fig3_concurrence}(b)-(c) illustrate the heteronuclear systems, where entanglement is intrinsically weaker due to asymmetric Larmor frequency. Even at low temperatures, the maximum concurrence is lower than that in homonuclear systems. Moreover, the critical value of $\omega_{\Sigma}/\boldsymbol{J}$, above which entanglement vanishes, shifts to higher values as $\omega_{\delta}/\boldsymbol{J}$ increases. This shift reflects the reduced overlap of spin populations in more asymmetric environments \cite{appelt2010paths,donovan2014heteronuclear}.

\subsubsection{Quantum Phase Transition in the Two-Spin-1/2 NMR System}

It is known that the non-analytic behavior of concurrence can serve as a signature of quantum phase transitions. Unlike classical phase transitions, quantum phase transitions occur only at very low temperatures ($T \rightarrow 0$), where quantum effects dominate over thermal fluctuations. Quantum phase transitions in two-qubit systems have already been explored in various settings in the literature \cite{arnesen2001natural,wang2001entanglement,zheng2015ansatz,di2024environment}. The two-spin system in our study can also present a quantum critical point as a signature of a phase transition \cite{arnesen2001natural,wang2001entanglement}.

The results in Fig.~\ref{Fig2_concurrence} and Fig.~\ref{Fig3_concurrence} highlight the appearance of quantum critical point induced by changes in $\omega_{\Sigma}/\boldsymbol{J}$ \cite{cheng2024quantum}. In Fig.~\ref{Fig2_concurrence}(a) and Fig.~\ref{Fig3_concurrence}(a), a distinct quantum critical point is evident at $\omega_{\Sigma}/\boldsymbol{J}=2$ . The analytical understanding of this observation can be derived from Eq.~(\ref{concurrence_simple}) in the low-temperature limit ($\beta \to \infty$). For $ \omega<\boldsymbol{J} $, $ {C} $ reaches 1, signifying a maximally entangled phase. Conversely, for $ \omega>\boldsymbol{J} $, $ {C} $ gives 0, indicating a transition to a non-entangled phase.   Therefore, the critical value of $ \omega $ at which the critical point occurs is given by $\omega = \boldsymbol{J}$. 
This condition coincides with the results of two-qubit phase transition in \cite{arnesen2001natural}.
This transition is marked by a sharp drop in concurrence at low temperatures (near zero), indicating a non-analytic change in the entanglement properties of the ground state. The transition separates a highly entangled phase from a phase where entanglement is zero. 
A similar quantum phase transition is shown in Fig.~\ref{Fig2_concurrence}(b), where the phase change is observed as the field changes from $\omega_{\Sigma}/\boldsymbol{J} = 2.3$ to $2.5$, and in Fig.~\ref{Fig2_concurrence}(c), where the field changes from $\omega_{\Sigma}/\boldsymbol{J} = 3.6$ to $3.8$. These shifts indicate that increasing the external magnetic field further suppresses entanglement.  Moreover, a larger $\omega_{\delta}$ leads to a higher critical magnetic field $\omega_{\Sigma}/\boldsymbol{J}$ for the transition between entangled and non-entangled phases. As the temperature rises, thermal fluctuations gradually diminish the entanglement, eventually driving the system into a thermally mixed state with vanishing entanglement above the $T_t$.

Generally speaking, the observed quantum critical point in this work arises from the interplay between spin-spin coupling and spin-field interactions \cite{osborne2002entanglement,chaboussant1998nuclear}. Physically, this transition represents a competition between the external magnetic field, which favors spin alignment and suppresses entanglement, and the exchange interaction strength, $\boldsymbol{J}$, which promotes anti-aligned, entangled spin configurations.

\subsection{NMR Signal Simulation and   Analysis of the Critical Point}

\begin{figure}[h]
\centering
  \includegraphics[width=\linewidth]{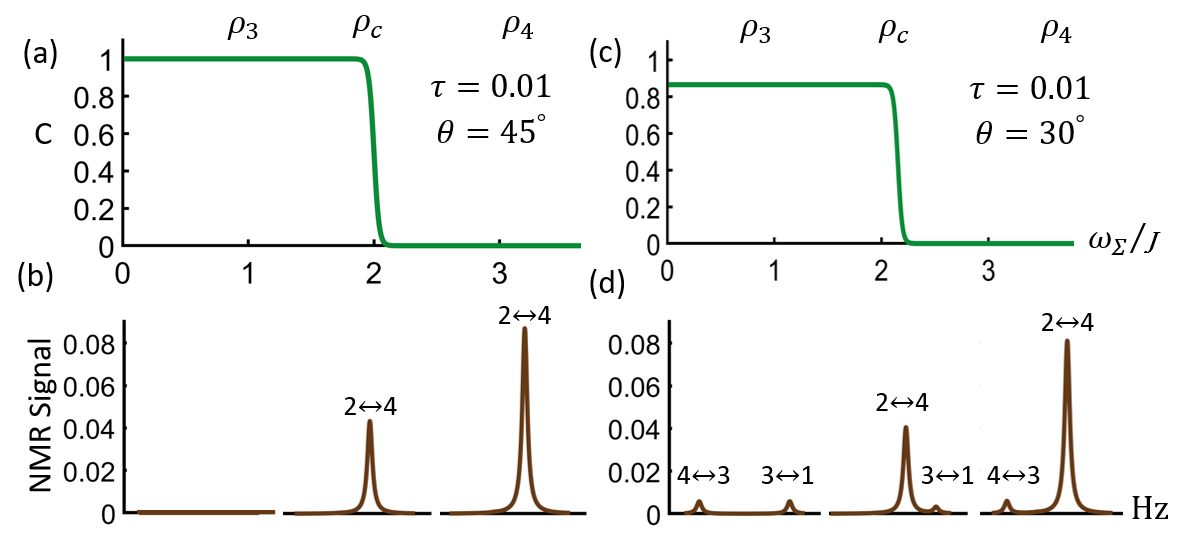}
\caption{Quantum signatures of ground-state transitions in (left-panel) homonuclear and (right-panel) heteronuclear spin systems. The concurrence $C$ sharply drops at the critical point at low temperature ($\tau = 0.01$). Bottom panels show simulated NMR spectra for mixing angle $\theta = 45^\circ$ (homonuclear) and $\theta = 30^\circ$ (heteronuclear), with flip angle $\varphi = 5^\circ$. }
\label{ground_state}
\end{figure}   

To analyze how ground-state degeneracy governs entanglement in NMR spin systems in the low-temperature limit ($\beta \to \infty$), we consider three representative scenarios based on the ordering of the two lowest energy levels. Figure~\ref{ground_state} summarizes these cases, combining concurrence and simulated spectral data across homonuclear and heteronuclear systems.
The first row shows the concurrence $C$ as a function of $\omega_\Sigma/\boldsymbol{J}$. The left column corresponds to a homonuclear system with $\theta = 45^\circ$, while the right column represents a heteronuclear system with $\theta = 30^\circ$ at $\tau = 0.01$. The second row displays the corresponding NMR spectra for each case, derived from transition amplitudes using the framework of Ref.\cite{vuichoud2015measuring} (see Appendix\ref{app1endix} for details). A small flip angle $\varphi = 5^\circ$ was used in all simulations to reflect the weak excitation conditions common in NMR.

In the first case, when $E_3$ is the lowest energy level, the system relaxes into the pure entangled state $|\phi_3\rangle$, leading to the thermal density matrix $\rho_3 = |\phi_3\rangle \langle \phi_3|$. Setting $p_3 = 1$ and all other populations to zero in Eq. (\ref{densityMatrix_G}) yields maximal entanglement ($C = 1$), as shown in Fig.\ref{ground_state}(a), left panel, for the homonuclear case. The corresponding spectrum in Fig.~\ref{ground_state}(b), is NMR silent, consistent with forbidden transitions due to the occupation of the singlet state \cite{mamone2020singlet}.
In contrast, the heteronuclear case (right panel) shows partial entanglement with $C \approx 0.86$. The NMR spectrum exhibits visible transitions from $|\phi_3\rangle$ to $|\phi_1\rangle$ and $|\phi_4\rangle$ ($3 \leftrightarrow 1$ and $4 \leftrightarrow 3$), confirming that the third eigenstate remains the most populated.

A critical transition occurs when the system reaches the degeneracy point, $E_3 = E_4$. Here, the thermal state reduces to 
\begin{equation}
  \rho_{c}=\frac{1}{2}(|\phi_3\rangle \langle \phi_3| + |\phi_4\rangle \langle \phi_4|),  
\end{equation}
an equally weighted mixture of two eigenstates. At this point, concurrence drops sharply, signaling a quantum critical point driven by the breakdown of quantum correlations in the degenerate ground state, as seen at the midpoint of Fig.~\ref{ground_state}(a)-(c). The corresponding spectrum for the homonuclear case shows only the transition $4 \leftrightarrow 2$, since the selection rules suppress other lines. The corresponding spectrum for the heteronuclear case shows balanced transitions such as $4 \leftrightarrow 2$ and $3 \leftrightarrow 1$, while other lines are suppressed, reflecting the symmetry in the eigenstate occupation.

In the third scenario, when $E_4$ becomes the lowest energy level ($p_4 = 1$), the system relaxes to the pure state $\rho_4 = |\phi_4\rangle \langle \phi_4|$, with $C$ dropping to zero, as seen in the far-right regions of Fig. \ref{ground_state}(a–b,c–d). In the homonuclear case, only the $2 \leftrightarrow 4$ transition appears in the NMR spectrum, while in the heteronuclear case, transitions involving $|\phi_4\rangle$, such as $4 \leftrightarrow 3$ and $4 \leftrightarrow 2$, dominate the spectrum. Transitions like $3 \leftrightarrow 1$ vanish due to the absence of a population in excited states, highlighting how the ground state structure shapes the spectral and quantum features.

Together, Fig.~\ref{ground_state} show how magnetic field tuning and spin asymmetry control the structure of thermal states, their quantum correlations, and their spectroscopic signatures in NMR. Understanding these signatures provides a robust path to engineer entangled and pure states in real NMR systems. Therefore, the transition can be inferred from the NMR signal of the system.

\subsection{Magnetic Field Variation in Homonuclear and Heteronuclear Systems}

The Hamiltonian given in Eq.~(\ref{HamiltonianTotal}) yields four distinct energy eigenstates, as outlined in Eq.~(\ref{EnergyLevel}). Using the field cycling technique \cite{carravetta2004beyond,miesel2006coherence,kimmich2004field,kiryutin2019transport}, these energy levels shift progressively as the external magnetic field increases, moving the system from a zero-field to a low-field and, ultimately, a high-field. 
Here, we explore the transition between strong and
weak coupling regimes facilitated by the magnetic field.

\subsubsection{Magnetic Field-Dependent in a Homonuclear System}

\begin{figure}[h]
\centering
  \includegraphics[width=\columnwidth]{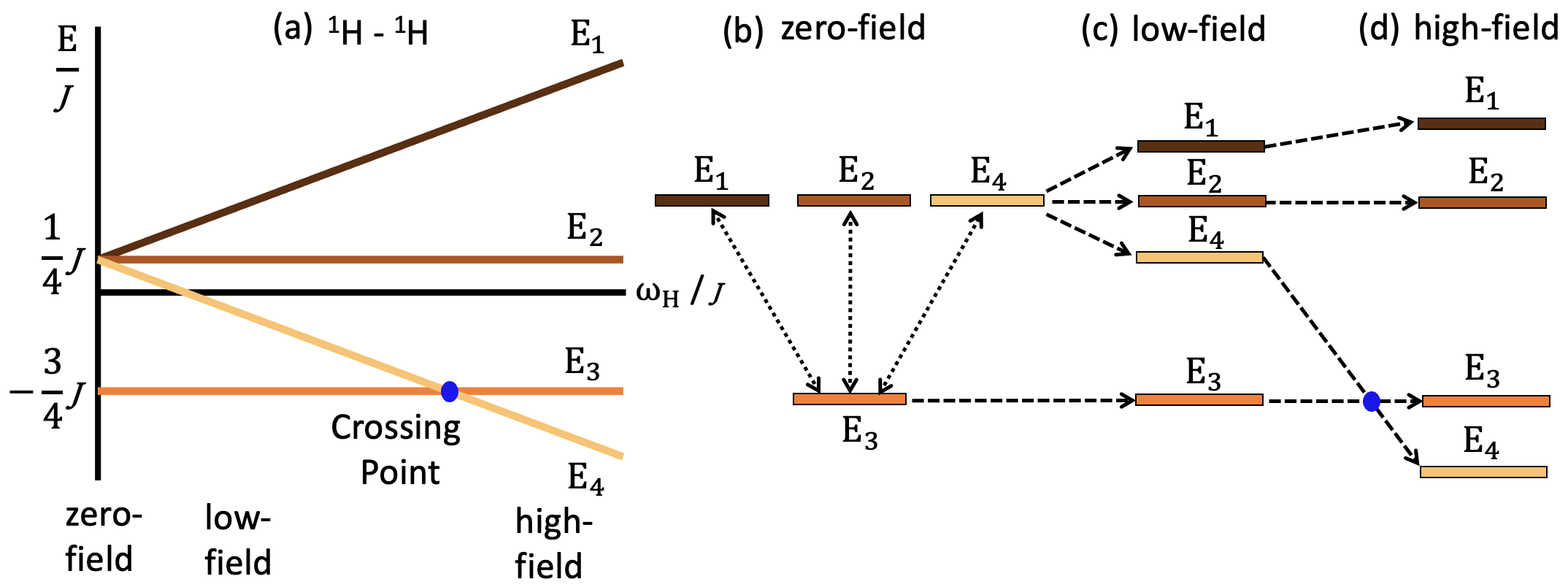}
	\caption{Field-dependent behavior of a homonuclear system (${}^{1}\text{H}-{}^{1}\text{H}$ system). The magnetic field strength increases from zero-field to low-field and finally reaches high-field.  A level crossing occurs between $\mathrm{E}_3$ and $\mathrm{E}_4$.}
	\label{Fig7_Field_dependent}
\end{figure}

We illustrate a field variation in a homonuclear system, where a two-spin-1/2 system changes from a zero-field to a low-field in a strong coupling regime and then a high-field in a weak coupling regime in Fig. \ref{Fig7_Field_dependent}. 
In the case of a homonuclear such as ${}^{1}\text{H}-{}^{1}\text{H}$ system, the energy level diagram is illustrated in Fig.~\ref{Fig7_Field_dependent}(a). For protons, the gyromagnetic ratio is denoted by $\gamma_{H}$, giving the Larmor frequencies for both spins as $\omega_1 = \omega_2 = \gamma_{H} B_0 = \omega_{H}$. Under the conditions $\omega_{\delta} = 0$ and  $\omega_{\Sigma} = 2\omega_{H}$, where 
Larmor frequencies range from zero to several \text{MHz}, and the $\boldsymbol{J}$-coupling is on the order of \text{Hz}. The resulting energy levels, from Eq.~(\ref{EnergyLevel}), are
\begin{equation}
\mathrm{E}_1 =\omega_{H}+\frac{1}{4}\boldsymbol{J}, \quad
\mathrm{E}_2=  \frac{1}{4}\boldsymbol{J}, \quad
\mathrm{E}_3 = - \frac{3}{4}\boldsymbol{J},\quad
\mathrm{E}_4 =-\omega_{H}+\frac{1}{4}\boldsymbol{J}. 
\end{equation}

This shows that the difference between $\mathrm{E}_2$ and $\mathrm{E}_3$ is determined by only the $\boldsymbol{J}$-coupling, which is usually small in organic NMR molecules [see Table~\ref{table1}]. In this setting $\omega_{\delta} = 0$, and the mixing angle is $\theta = \pi/4$.  Thus, the eigenstates match those given in Eqs.~(\ref{angular_T}) and (\ref{angular_S}).
In the absence of a magnetic field, as shown in Fig. \ref{Fig7_Field_dependent}(b), the Zeeman splitting vanishes, leading to degenerate triplet states with energies $\mathrm{E}_{1}=\mathrm{E}_{2}=\mathrm{E}_{4}=\boldsymbol{J}/4$, while the singlet state remains at $\mathrm{E}_{3}=-3\boldsymbol{J}/4$. 
The populations of these states are governed by thermal equilibrium. At low temperatures, the lower-energy singlet state is predominantly occupied, leading to strong entanglement. As the temperature rises, the population shifts into the triplet states, which reduces the overall entanglement. The density matrix reflects this shift, showing an increasing thermal population in the triplet manifold, which gradually diminishes the singlet-triplet coherence and the associated entanglement.

As the system moves from zero-field to a low-field regime, shown in Fig. \ref{Fig7_Field_dependent}(c), the introduction of a magnetic field lifts the triplet degeneracy. The Zeeman interaction causes a splitting of the energy levels, leading to the formation of distinct eigenstates. In the strong coupling regime, the entanglement starts decreasing, but it remains significant because $\boldsymbol{J}$-coupling is still stronger than the Zeeman effect.
However, the energy levels $ \mathrm{E}_2 $ and $ \mathrm{E}_3 $ remain unchanged, maintaining their positions as in the zero-field case. 
This results in the emergence of resonance transitions, as seen in the NMR spectrum in Fig.~\ref{Fig1_evergyLevel}(b), where $ D-\boldsymbol{J} = 0$, and $ D+\boldsymbol{J} = 2\boldsymbol{J}$. 
Moreover, transition from a zero-field to a low-field, the energy levels undergo significant changes, denoted as $\mathrm{E}^{zero}$ and $\mathrm{E}^{low}$, respectively. Specifically, the shift in energy for $\mathrm{E}^{zero}_{1}$ to $\mathrm{E}^{low}_{1}$ is $\Delta \mathrm{E} = {\omega_{\Sigma}}/{2} = \omega_{H}$. This positive shift signifies that the energy levels increase as the magnetic field strength is increased. Conversely, the energy transition from $\mathrm{E}^{zero}_{4}$ to $\mathrm{E}^{low}_{4}$ exhibits a decrease in energy as the magnetic field strength increases. This downward shift is quantified as  $\Delta \mathrm{E} = -{\omega_{\Sigma}}/{2} = -\omega_{H}$.

In the high-field regime, illustrated in Fig. \ref{Fig7_Field_dependent}(d), the Zeeman interaction dominates, and the energy levels align more closely with the individual spin projections along the field direction. This results in a spectrum characteristic of weak coupling, where the splittings are primarily determined by the Zeeman shifts and a reduced contribution from $\boldsymbol{J}$.  Thus, as the field increases further, entanglement eventually vanishes, leading to a non-entangled spin alignment in the high field. 
Figure~\ref{Fig7_Field_dependent}(c)-(d) illustrates the intersection of energy eigenstate $\mathrm{E}_{3}$ with $\mathrm{E}_{4}$ as the system transitions from low-field to high-field, introducing a quantum phase transition in the system at the crossing point. 

\subsubsection{Magnetic Field-Dependent in a Heteronuclear System}

\begin{figure}[h]
\centering
  \includegraphics[width=\columnwidth]{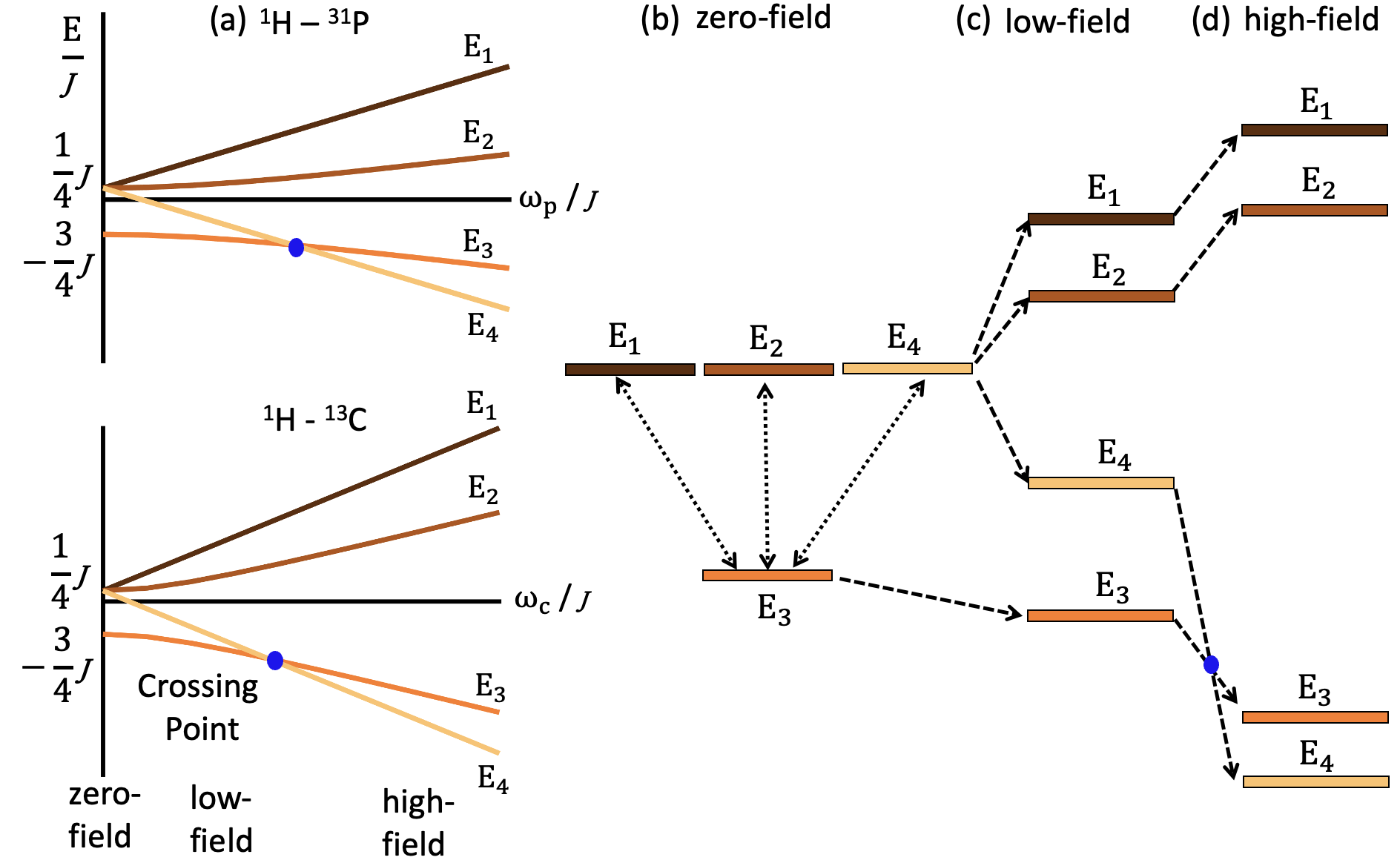}
	\caption{Field-dependent behavior of a heteronuclear system such as ${}^{1}\text{H}-{}^{31}\text{P}$ and ${}^{1}\text{H}-{}^{13}\text{C}$. The magnetic field strength increases from zero-field to low-field and finally reaches high-field. A level crossing occurs between $\mathrm{E}_3$ and $\mathrm{E}_4$}
	\label{Fig8_Field_dependent}
\end{figure}

The structure of heteronuclear systems differs from that of homonuclear systems. In these systems, spin couplings tend to be weaker, and the correlation between the two spins is generally lower. 
For example, for the heteronuclear interaction in ${}^{1}\text{H}-{}^{31}\text{P}$ system, the $\gamma_{H}$ is approximately $2.5$ times that of phosphorus  ($\gamma_p$), where $\gamma_{H} \approx 2.5\gamma_p$.
The substantial difference in Larmor frequencies of hydrogen ($\omega_{H}$) and phosphorus  ($\omega_p$) results in distinct energy levels, where $\omega_{H} \approx 2.5\omega_p$, as illustrated in Fig. \ref{Fig8_Field_dependent}(a). 
In this scenario, we observe $\omega_{\delta} \approx 1.5 \omega_p$ and $\omega_{\Sigma} \approx 3.5 \omega_p$  in the energy level equations due to the significant differences in Larmor frequencies. Additionally,  
$D \simeq \boldsymbol{J} \sqrt{9(\omega_p/\boldsymbol{J})^2/4+1}$.  By substituting these conditions into Eq. (\ref{EnergyLevel}), we have the energy levels as
\begin{equation}
 \begin{aligned}
&\mathrm{E}_1=\frac{7}{4} {\omega_p}+\frac{1}{4}\boldsymbol{J},  \quad
\mathrm{E}_{2}=\frac{\boldsymbol{J}}{2} \sqrt{\frac{9}{4}(\frac{\omega_p}{\boldsymbol{J}})^2+1}-\frac{1}{4}\boldsymbol{J}, \quad
\\
&\mathrm{E}_3=-\frac{\boldsymbol{J}}{2} \sqrt{\frac{9}{4}(\frac{\omega_p}{\boldsymbol{J}})^2+1}-\frac{1}{4}\boldsymbol{J}, \quad 
\mathrm{E}_4=-\frac{7}{4}{\omega_p}+\frac{1}{4}\boldsymbol{J}.
\end{aligned} 
\end{equation}

In this setting, transition from a zero-field to a low-field, the energy difference for $\mathrm{E}^{zero}_{1}$ to $\mathrm{E}^{low}_{1}$ is $\Delta \mathrm{E}_1 = \mathrm{E}^{low}_{1} - \mathrm{E}^{zero}_{1} = {7\omega_{p}}/{4}$.  Similarly, the energy difference for $\Delta\mathrm{E}_2$ is follows a different functional form $\Delta \mathrm{E}_2 = {\boldsymbol{J}} \sqrt{9(\omega_p/\boldsymbol{J})^2/4+1}/2 -  \boldsymbol{J}/2$ but remains lower than $\Delta \mathrm{E}_1$  indicating an asymmetry in the energy level shifts under a low-field regime. Conversely, the energy transition from $\mathrm{E}^{zero}_{4}$ to $\mathrm{E}^{low}_{4}$ is quantified as  $\Delta \mathrm{E}_4 = -{7\omega_{p}}/{4}$.

We can similar analysis for the ${}^{1}\text{H}-{}^{13}\text{C}$ system. In this case, the $\gamma_{H}$ is approximately four times that of carbon ($\gamma_c$), with $\gamma_{H} \approx 4\gamma_c$, where $\omega_{H} \approx 4\omega_c$, is illustrated in Fig. \ref{Fig8_Field_dependent}(a).
The distinct values for $\omega_{\delta}$ and $\omega_{\Sigma}$ emerge in the energy level equations, namely $\omega_{\delta} \approx 3 \omega_c$ and $\omega_{\Sigma} \approx 5 \omega_c$, due to the substantial differences in Larmor frequencies. Additionally, the expression for $D$ can be approximated as $D \simeq \boldsymbol{J} \sqrt{9(\omega_c/\boldsymbol{J})^2+1}$. Substituting these conditions into Eq. (\ref{EnergyLevel}), the resulting energy levels are expressed as
\begin{equation}
 \begin{aligned}
&\mathrm{E}_1=\frac{5}{2} {\omega_c}+\frac{1}{4}\boldsymbol{J},  \quad 
\mathrm{E}_{2}=\frac{\boldsymbol{J}}{2} \sqrt{9(\frac{\omega_c}{\boldsymbol{J}})^2+1}-\frac{1}{4}\boldsymbol{J}, \quad 
\\
&\mathrm{E}_3=-\frac{\boldsymbol{J}}{2} \sqrt{9(\frac{\omega_c}{\boldsymbol{J}})^2+1}-\frac{1}{4}\boldsymbol{J}, \quad 
\mathrm{E}_4=-\frac{5}{2} {\omega_c}+\frac{1}{4}\boldsymbol{J}.
\end{aligned} 
\end{equation}

Figure \ref{Fig8_Field_dependent}(b)-(d) illustrates the field-dependent evolution of a heteronuclear system as it transitions from zero-field to low-field and then to high-field. 
In the absence of a magnetic field (zero-field), shown in Fig.~\ref{Fig8_Field_dependent}(b), the Zeeman splitting is absent, and the system exhibits a degenerate triplet state as mentioned in Fig.~\ref{Fig7_Field_dependent}(b).
As the system moves into the low-field regime, depicted in Fig.~\ref{Fig8_Field_dependent}(c), the Zeeman interaction starts to lift the triplet degeneracy. The heteronuclear nature of the system leads to an asymmetric splitting pattern, where the energy levels shift unevenly. 
In the high-field regime, shown in Fig.~\ref{Fig8_Field_dependent}(d), the Zeeman interaction dominates, aligning the energy levels with the individual spin projections along the external field. Unlike in the homonuclear case, the heteronuclear system exhibits an additional asymmetry, where the energy levels $\mathrm{E}_1$ and $\mathrm{E}_2$ become more separated due to the Zeeman splitting, while $\mathrm{E}_3$ and $\mathrm{E}_4$ align differently as the system moves towards weak coupling. Similar to the homonuclear system, a quantum pcritical point emerges as the field strength increases, occurring at the crossing point where the energy eigenstates $\mathrm{E}_3$ and $\mathrm{E}_4$ intersect.

\subsubsection{General Criteria for the quantum critical point in the Two-Spin-1/2 NMR System}

The quantum critical point in two-spin systems can emerge under certain conditions, particularly in models like the isotropic Heisenberg Hamiltonian as ~\cite{arnesen2001natural, nielsen2000quantum}
\begin{equation}\label{Heisenberg}
\mathcal{H} = \mu_B\vec{B}(\sigma_{1z} + \sigma_{2z})+\boldsymbol{J}\, \vec{\sigma}_1 \cdot \vec{\sigma}_2. 
\end{equation}
This Hamiltonian is a specific case of our general form in Eq.~(\ref{Hamiltonian_G}). However, the emergence of a quantum critical point is not exclusive to this particular model but can arise more broadly depending on the system's parameters and symmetries.

To clarify, we investigate a specific two-spin system that can shed light on the spin entanglement structure. To this aim,
we consider the entanglement in the hyperfine structure of the hydrogen atom, analyzed by Maleki \textit{et al.}~\cite{maleki2021naturalennett}. In this system, the Hamiltonian includes a spin-spin interaction between the electron and proton, as well as a Zeeman interaction acting only on the electron as
\begin{equation} \label{hyperfine_electron}
\mathcal{H}_{\text{HF}} = \mu_B B \vec{\sigma}_e  + A\, \vec{\sigma}_e \cdot \vec{\sigma}_p.
\end{equation}

To establish a clear connection between our general two-spin-1/2 Hamiltonian and the hyperfine spin-spin coupling model used in atomic hydrogen systems, we begin by rewriting our Hamiltonian Eq. (\ref{Hamiltonian_G}) in terms of Pauli operators as 
\begin{equation}
\mathcal{H}= \frac{\omega_1}{2} \sigma_{1z} + \frac{\omega_2}{2} \sigma_{2z} + \frac{\boldsymbol{J}}{4} \vec{\sigma}_1 \cdot \vec{\sigma}_2.
\end{equation}

To make a direct correspondence with the hyperfine model, we observe that the spin-spin interaction term $ {\boldsymbol{J}} \vec{\sigma}_1 \cdot \vec{\sigma}_2/{4} $ in our Hamiltonian becomes identical to the hyperfine coupling term $ A\, \vec{\sigma}_e \cdot \vec{\sigma}_p $ if we set $ \boldsymbol{J} = 4A $. Moreover, the Zeeman interaction $ \mu_B \vec{\sigma}_e \cdot \vec{B} $ in Eq. (\ref{hyperfine_electron}) acts exclusively on the electron spin, whereas our general Hamiltonian includes Zeeman terms for both spins. 
To recover the hyperfine form, we set the Zeeman interaction of the proton to be zero by setting $ \omega_2 = 0$ (neglecting the proton’s Zeeman term due to its small magnetic moment). With these substitutions, our Hamiltonian reduces to
\begin{equation}
\mathcal{H} = \frac{\omega_1}{2} \sigma_{1z} + A\, \vec{\sigma}_1 \cdot \vec{\sigma}_2,
\end{equation}
which corresponds to the effective spin Hamiltonian used to describe hyperfine coupling in atomic hydrogen. This demonstrates that the hyperfine model represents a limiting case of our general two-spin-1/2 system, characterized by isotropic spin-spin coupling and a Zeeman interaction that dominantly affects only the electron.  
This shows that the hyperfine model is a limiting case of the general two-spin-1/2 Hamiltonian, featuring isotropic coupling and an asymmetric Zeeman term.
However, in this model, no level crossing occurs. This raises the question: what is the general criterion for observing a quantum critical point in two-spin systems?

The emergence of a level crossing between energy states $E_3$ and $E_4$ serves as a key signature of a quantum critical point. This crossing occurs when the following condition is 
\begin{equation}
\boldsymbol{J} = \frac{\omega^2_{\Sigma}-\omega^2_{\delta}}{2\omega_{\Sigma}}.
\label{crossingPoint}
\end{equation}
This equation provides a unified framework for identifying the crossing point in both homonuclear and heteronuclear systems.
Alternatively, this expression can also be rewritten as
\begin{equation}
\boldsymbol{J} = 2\frac{\omega_1\omega_2}{\omega_1 + \omega_2}.
\label{crossingPoint_2}
\end{equation}

In the homonuclear case, 
i.e., $\omega_1 = \omega_2 = \omega$, the quantum critical point occurs at $J=\omega$, which agrees with the observation for the isotropic Heisenberg model of Eq. (\ref{Heisenberg}) reported in \cite{arnesen2001natural}.
To be more specific, 
in a homonuclear ${}^{1}\text{H}$--${}^{1}\text{H}$ spin pair, symmetry in the gyromagnetic ratios simplifies this condition to $\boldsymbol{J} = \omega_{\Sigma}/2 = \omega_{H} = \gamma_{H}B$, yielding the critical magnetic field $B = \boldsymbol{J}/\gamma_{H}$ in proton NMR.
When applied this crossing condition to the hyperfine structure,  Eq. (\ref{crossingPoint_2}) cannot be satisfied for $\omega_2 = 0$. As a result, no level crossing occurs, and therefore, no quantum critical point can be realized in this model.

For heteronuclear systems, the crossing condition provides useful estimates for critical magnetic fields. In the ${}^{1}\text{H}$--${}^{13}\text{C}$ system, it simplifies to $\boldsymbol{J} \approx 0.4\, \omega_{H}$, while for a ${}^{1}\text{H}$--${}^{31}\text{P}$ pair, it becomes $\boldsymbol{J} \approx 0.57\, \omega_{H}$. These relations yield corresponding critical field values of $B = 2.5\, \boldsymbol{J}/\gamma_{H}$ in carbon NMR and $B = 1.75\, \boldsymbol{J}/\gamma_{H}$ in phosphorus NMR. These critical fields define the parameter regimes where level crossings can occur in coupled spin systems. In contrast, the hyperfine interaction in the hydrogen atom does not exhibit such a transition due to the dominant Zeeman interaction acting only on the electron, which prevents any crossing between the energy levels $E_3$ and $E_4$.

To further examine the generality of this criterion, we consider the positronium system. The positronium Hamiltonian is given by~\cite{sauder1967zeeman,mogensen2012positron,pachucki1997effective}
\begin{equation}
\mathcal{H}_{\text{Ps}} = \mu_B B\, (\sigma_{1z} - \sigma_{2z}) + \mathcal{A'}\, \boldsymbol{\sigma}_1 \cdot \boldsymbol{\sigma}_2,
\end{equation}
where $\mathcal{A'}$ is the spin-spin (hyperfine) coupling coefficient, and $\mu_B B$ represents the Zeeman interaction under an external magnetic field. This Hamiltonian can be viewed as a specific case of our general form Eq. (\ref{Hamiltonian_G}), as
\begin{equation} \label{Hamiltonian_Re}
\mathcal{H} = \frac{\omega_\Sigma}{4}(\sigma_{1z} + \sigma_{2z}) + \frac{\omega_\delta}{4}(\sigma_{1z} - \sigma_{2z}) + \frac{\boldsymbol{J}}{4} \, \boldsymbol{\sigma}_1 \cdot \boldsymbol{\sigma}_2.
\end{equation}

In the positronium system, the electron and positron possess equal and opposite magnetic moments, leading to $\omega_\Sigma = 0$. This eliminates the symmetric Zeeman term $(\sigma_{1z} + \sigma_{2z})$, reducing the Hamiltonian to
\begin{equation}
\mathcal{H} = \frac{\omega_\delta}{4}(\sigma_{1z} - \sigma_{2z}) + \frac{J}{4} \, \boldsymbol{\sigma}_1 \cdot \boldsymbol{\sigma}_2,
\end{equation}
with the identifications $\omega_\delta = 4 \mu_B B$ and $J = 4 \mathcal{A'}$. This mapping shows that positronium is a particular realization of the two-spin-1/2 model under symmetric detuning and rescaled coupling strength.
However, since $J = 4 \mathcal{A'} > 0$, the quantum critical point in Eq.~(\ref{crossingPoint}) cannot be fulfilled. Thus, positronium, like the hydrogen hyperfine system, does not exhibit a quantum phase transition due to the absence of level crossing between $E_3$ and $E_4$.

\subsection{Experimental Quantification of Entanglement Using NMR Polarization }

Despite the importance of entanglement in quantum sciences, the experimental quantification of entanglement measures is quite challenging. This is due to the fact that entanglement measures are generally not related to an observable operator. Most methods rely on full quantum state tomography, which is resource-intensive and experimentally demanding \cite{lu2016tomography,lvovsky2009continuous,gross2010quantum,xin2017quantum}. Here, we present an experimentally viable approach that reconstructs the degree of entanglement of a two-spin quantum system using three measurable NMR observables: the individual Zeeman polarization term $ \mathrm{Tr}(\sigma_{1z} \rho) $, $ \mathrm{Tr}(\sigma_{2z} \rho) $, and the longitudinal two-spin order measurement $\mathrm{Tr}(\sigma_{1z} \sigma_{2z}\rho)$. 
These observables can be extracted from standard one-dimensional NMR experiments via customized pulse sequences. Our framework bridges the gap between abstract quantum information quantities and experimentally accessible data, enabling robust characterization of quantum correlations.

To analyze the concurrence and expectation values in a two-spin system, we evaluate the expectation value of $\sigma_{1z}$, with the density matrix $\rho$ from Eq.~(\ref{densityMatrix_G}), projected in $I_{1z}$ basis. The general expression for the expectation value of $\sigma_{1z}$ is given by ~\cite{vuichoud2015measuring}
\begin{equation}
    \mathrm{P}_{1z} = \mathrm{Tr}(\sigma_{1z} \rho) = \sum_i p_i \langle \phi_i | \sigma_{1z} | \phi_i \rangle= p_1 - p_4 + (p_2 - p_3) \cos 2\theta.
\end{equation} 
Similarly, the observable projection of the second spin is given by
\begin{equation}
    \mathrm{P}_{2z} = \mathrm{Tr}(\sigma_{2z} \rho) = \sum_i p_i \langle \phi_i | \sigma_{2z} | \phi_i \rangle = p_1 - p_4 + (p_3 - p_2) \cos 2\theta.
\end{equation}
The longitudinal two-spin expectation can be expressed as
\begin{equation}
\mathrm{P}_{1z,2z} = \mathrm{Tr}(\sigma_{1z} \sigma_{2z}\rho) = \sum_i p_i \langle \phi_i |\sigma_{1z} \sigma_{2z} | \phi_i \rangle = p_1 + p_4 - (p_3 + p_2).
\end{equation}
Assuming $\cos 2\theta \neq0$, these three observables allow the full reconstruction of the populations as
\begin{equation}
\begin{aligned}
p_1 &= \frac{1}{4} (1 + \mathrm{P}_{1z} + \mathrm{P}_{2z} + \mathrm{P}_{1z,2z}), \quad
p_2 = \frac{1}{4} (1 - \mathrm{P}_{1z,2z} + \frac{\mathrm{P}_{1z} - \mathrm{P}_{2z}}{\cos 2\theta}), \\
p_3 &= \frac{1}{4} (1 - \mathrm{P}_{1z,2z} + \frac{\mathrm{P}_{2z} - \mathrm{P}_{1z}}{\cos 2\theta}), \quad
p_4 = \frac{1}{4} (1 - \mathrm{P}_{1z} - \mathrm{P}_{2z} + \mathrm{P}_{1z,2z}).
\end{aligned}
\label{eq_populations_corrected}
\end{equation}
Substituting the above terms into the concurrence $C$ Eq. (\ref{concurrence_G})
provides a complete expression for entanglement in terms of NMR observables as 
\begin{equation}
C = \frac{1}{2} \max\{ 0,\ | \mathrm{P}_{1z} - \mathrm{P}_{2z} | |\tan 2\theta| 
- \sqrt{1 + \mathrm{P}^2_{1z,2z} - ( \mathrm{P}_{1z} + \mathrm{P}_{2z} )^2 } \}.
\end{equation}
The polarization terms can be extracted from expectation values, while the mixing angle factor $\tan 2\theta$ can be inferred from the NMR spectral pattern shown in Fig.~\ref{Fig1_evergyLevel}.

These results demonstrate that concurrence which is typically regarded as mathematical quantifier of quantum entanglement, can in fact be experimentally determined using standard NMR polarization measurements. This approach provides a practical framework for analyzing entanglement and quantum features in spin-1/2 systems, thus offering a powerful tool set for quantum information benchmarking in real-world NMR experiments.

\section{Discussions}

Previous studies have investigated quantum entanglement in spin-1/2 dimers, focusing on the effects of temperature and magnetic field on entanglement degree of the systems \cite{arnesen2001natural,wang2001entanglement,guo2003thermal,asoudeh2005thermal,vcenvcarikova2020unconventional,adamyan2020quantum}. Thermal entanglement in Heisenberg models and entanglement-assisted quantum critical point were considered in \cite{arnesen2001natural,wang2001entanglement}. The dynamical evolution of entanglement in a Heisenberg model has also been explored in the literature \cite{abliz2006entanglement,wang2019entanglement}. More recent works, such as \cite{adamyan2020quantum,vcenvcarikova2020unconventional}, extended these studies to systems with  mixed spin-(1/2, 1) systems, analyzing both ground-state entanglement \cite{adamyan2020quantum,vcenvcarikova2020unconventional} and thermal entanglement \cite{vcenvcarikova2020unconventional}.
The impact of magnetic field inhomogeneity on thermal entanglement has also been investigated in \cite{asoudeh2005thermal}, where an isotropic Heisenberg two-spin system was subjected to nonuniform Zeeman fields. It was shown that the entanglement can also be modified by variations in the anisotropy parameter in the system, resulting in an altered threshold temperature in this system \cite{guo2003thermal}.

Despite their theoretical value, many models studied in the literature remain abstract and not directly connected to experimentally measurable parameters in NMR. This fact limits their applicability to real NMR systems and emerging quantum sensing platforms. Our work addresses this gap by formulating a spin system explicitly grounded in NMR-relevant parameters and settings. The two-spin-1/2 configuration, which can be directly realized in both liquid- and solid-state NMR experiments, offers a natural platform to connect theoretical quantum properties with observable quantities. In this study, we exploit this connection by analyzing thermal entanglement in spin pairs such as ${}^{1}\text{H}$–${}^{1}\text{H}$, ${}^{1}\text{H}$–${}^{31}\text{P}$, and ${}^{1}\text{H}$–${}^{13}\text{C}$ under realistic field and coupling conditions. By doing so, we demonstrate how quantum information concepts like concurrence can be quantitatively assessed, and potentially manipulated, in practical NMR experiments.
By translating quantum information insights into a language that is more intuitive for both experimental and theoretical chemists, this work also contributes to addressing the challenge highlighted in recent literature that emphasizes the need for more chemically intuitive formulations of quantum information concepts to foster cross-disciplinary understanding and research \cite{scholes2025quantum, aliverti2024can,wu2024foundations,zhang2024molecular,delgado2025quantum}.

\section{Conclusion}

In this work, we presented a comprehensive analytical and numerical framework for investigating quantum entanglement in two-spin-1/2 NMR systems under thermal and magnetic field conditions. By deriving closed-form expressions for concurrence, we systematically analyzed how this quantum feature is governed by  temperature, spin-spin coupling strength, and magnetic field. The resulting analytical expression for concurrence enables precise identification of the threshold temperature beyond which entanglement disappears. Our framework encompasses both homonuclear and heteronuclear spin-pair configurations, allowing exploration of a wide range of realistic NMR systems. At absolute zero temperature, we demonstrated that the system may exhibit a quantum critical point, marked by a non-analytic change in concurrence. To understand the origin of this critical behavior,
we introduced a general analytical criterion for energy level crossing, which serves as a tool for identifying quantum phase transitions. This criterion was applied to a variety of representative systems, including homonuclear and heteronuclear spin pairs as well as hyperfine and positronium configurations, providing a unified understanding of the emergence of the quantum criticality in different physical models. Furthermore, we established a direct correspondence between quantum entanglement  and experimentally accessible NMR observables. This enables the reconstruction of entanglement using standard polarization measurements. By bridging theoretical quantum quantifiers with practical NMR techniques, our work enables experimental quantification and control of quantum correlations in realistic NMR systems.
These findings offer insights into the control of spin entanglement, with implications for quantum sensing, spin-based thermometry, and entanglement-assisted NMR. This framework provides a powerful toolset for exploring non-classical features in experimentally accessible platforms and contributes to the integration of quantum information theory with practical NMR technology.

\section*{Conflicts of interest
}
There are no conflicts to declare.

\section*{Acknowledgements}
This project was supported by the Cancer Prevention and Research Institute of Texas (CPRIT) under Grant No. RR240015.

\section*{Appendix}

\subsection*{Transition Amplitudes} \label{app1endix}

The transition amplitudes ${i \leftrightarrow j}$ characterize the coherent oscillations between the energy levels of a two-spin-1/2 system under the influence of an RF pulse. These expressions depend on the flip angle $\varphi$, the mixing angle $\theta$, and the population differences between the levels $p_i - p_j$. Below, we analyze each transition and its simplified cases. These general forms account for all population differences and interference between transitions via $p_1$ through $p_4$,  are  given as ~\cite{vuichoud2015measuring}
{\small
\begin{align}
{4 \leftrightarrow 3} &= -\frac{1}{2} \sin\varphi \left[
    \sin^2( \frac{\varphi}{2})(1 - \sin 2\theta)(p_3 - p_1) \right. \nonumber \\
&\quad \left. - \sin^2( \frac{\varphi}{2}) \cos^2(2\theta)(p_3 - p_2)
    + \cos^2(\frac{\varphi}{2})(1 - \sin 2\theta)(p_4 - p_3)
\right], \nonumber\\
{2 \leftrightarrow 1} &= -\frac{1}{2} \sin\varphi \left[
    \cos^2(\frac{\varphi}{2})(1 + \sin 2\theta)(p_2 - p_1) \right.\nonumber\\
&\quad \left. - \sin^2(\frac{\varphi}{2}) \cos^2(2\theta)(p_3 - p_2)
    + \sin^2(\frac{\varphi}{2})(1 + \sin 2\theta)(p_4 - p_2)
\right], \nonumber\\
{4 \leftrightarrow 2} &= -\frac{1}{2} \sin\varphi \left[
    \sin^2(\frac{\varphi}{2})(1 + \sin 2\theta)(p_2 - p_1) \right. \nonumber\\
&\quad \left. + \sin^2(\frac{\varphi}{2}) \cos^2(2\theta)(p_3 - p_2)
    + \cos^2(\frac{\varphi}{2})(1 + \sin 2\theta)(p_4 - p_2)
\right], \nonumber\\
{3 \leftrightarrow 1} &= -\frac{1}{2} \sin\varphi \left[
    \cos^2(\frac{\varphi}{2})(1 - \sin 2\theta)(p_3 - p_1) \right.\nonumber \\
&\quad \left. + \sin^2(\frac{\varphi}{2}) \cos^2(2\theta)(p_3 - p_2)
    + \sin^2(\frac{\varphi}{2})(1 - \sin 2\theta)(p_4 - p_3)
\right].
\end{align}}

As the first setting, we consider $p_1 = p_2 = p_4 = 0$, $p_3 = 1$.  The details of the transition amplitudes are
\begin{align}
{4 \leftrightarrow 3} &= -\frac{1}{2} \sin\varphi \left[
    \sin^2(\frac{\varphi}{2})(1 - \sin 2\theta)
    - \sin^2(\frac{\varphi}{2})\cos^2(2\theta) \right. \nonumber\\
&\quad \left. - \cos^2(\frac{\varphi}{2})(1 - \sin 2\theta) \right],\nonumber \\
{2 \leftrightarrow 1} &= \frac{1}{2} \sin\varphi  \sin^2(\frac{\varphi}{2}) \cos^2(2\theta), \nonumber\\
{4 \leftrightarrow 2} &= -\frac{1}{2} \sin\varphi \sin^2(\frac{\varphi}{2}) \cos^2(2\theta), \nonumber\\
{3 \leftrightarrow 1} &= -\frac{1}{2} \sin\varphi \left[
    \cos^2(\frac{\varphi}{2})(1 - \sin 2\theta)
    + \sin^2(\frac{\varphi}{2})\cos^2(2\theta) \right. \nonumber\\
&\quad \left. - \sin^2(\frac{\varphi}{2})(1 - \sin 2\theta) \right].
\end{align}

For the second setting, we have  $p_1 = p_2 = 0$, $p_3 = p_4 = 1/2$. The details of the transition amplitudes are
\begin{align}
{4 \leftrightarrow 3} &= -\frac{1}{4} \sin\varphi \left[
    \sin^2(\frac{\varphi}{2})(1 - \sin 2\theta)
    - \sin^2(\frac{\varphi}{2})\cos^2(2\theta) \right], \nonumber\\
{2 \leftrightarrow 1} &= \frac{1}{4} \sin\varphi \left[
    \sin^2(\frac{\varphi}{2})\cos^2(2\theta)
    - \sin^2(\frac{\varphi}{2})(1 + \sin 2\theta) \right], \nonumber\\
{4 \leftrightarrow 2} &= -\frac{1}{4} \sin\varphi \left[
    \sin^2(\frac{\varphi}{2})\cos^2(2\theta)
    + \cos^2(\frac{\varphi}{2})(1 + \sin 2\theta) \right],\nonumber \\
{3 \leftrightarrow 1} &= -\frac{1}{4} \sin\varphi \left[
    \cos^2(\frac{\varphi}{2})(1 - \sin 2\theta)
    + \sin^2(\frac{\varphi}{2})\cos^2(2\theta) \right].
\end{align}

As the third setting, we set $p_1 = p_2 = p_3 = 0$, $p_4 = 1$.
 The transitions primarily occur from level 4 downward, and each path is modulated differently depending on whether cosine or sine powers dominate. The details of the transition amplitudes are 
\begin{align}
{4 \leftrightarrow 3} &= -\frac{1}{2} \sin\varphi  \cos^2(\frac{\varphi}{2})(1 - \sin 2\theta), \nonumber\\
{2 \leftrightarrow 1} &= -\frac{1}{2} \sin\varphi  \sin^2(\frac{\varphi}{2})(1 + \sin 2\theta), \nonumber\\
{4 \leftrightarrow 2} &= -\frac{1}{2} \sin\varphi \cos^2(\frac{\varphi}{2})(1 + \sin 2\theta),\nonumber\\
{3 \leftrightarrow 1} &= -\frac{1}{2} \sin\varphi  \sin^2(\frac{\varphi}{2})(1 - \sin 2\theta).
\end{align}

\renewcommand\refname{References}

\bibliography{rsc} 
\bibliographystyle{rsc} 

\end{document}